# Synchronisation, desynchronisation and intermediate regime of breathing solitons and soliton molecules in a laser cavity

Xiuqi Wu[1], Junsong Peng[1,2,3,*], Sonia Boscolo[4], Christophe Finot[5], and Heping Zeng[1,5,6,**]

[1]State Key Laboratory of Precision Spectroscopy, East China Normal University, Shanghai 200062, China

[2]Collaborative Innovation Center of Extreme Optics, Shanxi University, Taiyuan, Shanxi 030006, China

[3]Chongqing Key Laboratory of Precision Optics, Chongqing Institute of East China Normal University, Chongqing 401120, China

[4]Aston Institute of Photonic Technologies, Aston University, Birmingham B4 7ET, United Kingdom

[5]Laboratoire Interdisciplinaire Carnot de Bourgogne, UMR 6303 CNRS – Université de Bourgogne Franche-Comté, F-21078 Dijon Cedex, France

[6]Shanghai Research Center for Quantum Sciences, Shanghai 201315, China

*jspeng@lps.ecnu.edu.cn ** hpzeng@phy.ecnu.edu.cn

Abstract: We report on the experimental and numerical observations of synchronisation and desynchronisation of bound states of multiple breathing solitons (breathing soliton molecules) in an ultrafast fibre laser. In the desynchronisation regime, although the breather molecules as wholes are not synchronised to the cavity, the individual breathers within a molecule are synchronised to each other with a delay (lag synchronisation). An intermediate regime between the synchronisation and desynchronisation phases is also observed, featuring self-modulation of the synchronised state. This regime may also occur in other systems displaying synchronisation. Breathing soliton molecules in a laser cavity open new avenues for the study of nonlinear synchronisation dynamics.

First recognised in 1665 by Christiaan Huygens in a letter to his father [1], synchronisation, i.e., the ability of coupled oscillators to lock to a common frequency, is a general and ubiquitous feature of nature, occurring for biological clocks, chemical reactions, mechanical or electrical oscillators, and lasers to mention a few well-known examples [2]. Apart from being of significant interest in fundamental science, synchronisation phenomena also find a wide range of practical applications. For instance, in the field of optics, the synchronisation of multiple micro-resonators can break the power limitation of a single micro-resonator [3-5]. Breathing solitons, manifesting themselves as localised temporal or spatial structures that exhibit periodic oscillatory behaviour, are fundamental modes of many nonlinear physical systems and relate to a wide range of important nonlinear dynamics. In optics, initially studied in single-pass fibre systems [6, 7], the breathing soliton concept has been extended to passive Kerr cavities and micro-resonators [8-11] as well as to ultrafast fibre lasers [12-24]. The studies in [9, 14] have shown that cavity-based optical systems can support breather oscillations that are sub-harmonically entrained to the cavity roundtrip time. This sub-harmonic entrainment of breathers, which is a generalised form of synchronisation wherein a harmonic of the breathing frequency $f_b$ synchronises with the cavity frequency $f_r$, results from the competition between the two intrinsic frequencies to the system $f_r$ and $f_b$. In [21], we have reported frequency locking at Farey fractions of a breather fibre laser by demonstrating for the first time that the winding numbers $f_b/f_r$ show the hierarchy of a Farey tree and the structure of a devil's staircase, in accordance with the predictions from the theory of nonlinear systems with two competing frequencies [25].

These breather synchronisation studies pertain to the dynamics of single breathers. In like manner to their stationary counterparts [26-31], multiple interacting breathing solitons in dissipative systems can also organise themselves into molecule-like bound states [10, 13, 22]. It remains elusive whether breather molecular complexes also display synchronisation. Furthermore, while single breather oscillations in a laser represent a convenient nonlinear dynamical system to study two-frequency interactions [21], multi-breather complexes add new degrees of freedom into the system – the breathing frequencies of the elementary constituents - thereby opening the possibility to study the dynamics of nonlinear systems with three or more interacting frequencies, which is an important topic in nonlinear science [32-35].

In this Letter, we present the results of a further study showing that breather molecules can synchronise to a laser cavity, featuring a breathing frequency that equals a subharmonic of the cavity fundamental frequency. In the desynchronised phase, while the breather molecule as whole is not synchronised to the cavity, lag synchronisation among the constituent breathers is observed. An intermediate regime between the synchronised and desynchronised phases is also observed, featuring a subharmonic breathing frequency with non-subharmonic sidebands. Direct synchronisation-desynchronisation transitions without such a regime occur when the intracavity loss is increased. The transitions among these different phases are realised by tuning a single control parameter (the pump current). These experimental findings are confirmed by numerical simulations of a lumped laser model.

The system that we studied is a standard mode-locked fibre laser (Fig. 1). The gain medium is a 1.24 m-long erbium-doped fibre. Other fibres in the cavity are single-mode fibres. The group-velocity dispersion (GVD) parameters of the two fibre types are 61.2 and –18 $ps^2$/km, respectively, resulting in a slightly anomalous net cavity dispersion (–0.0078 $ps^2$). The laser has a repetition frequency of $f_r$ = 33.39 MHz. Mode locking is realised through an effective saturable absorber by the nonlinear polarisation evolution (NPE) effect [36]. The transfer function of NPE is controlled via three wave plates working together with a polarisation beam splitter which also serves as the laser output port. The pump strength is the key parameter of the system: its tuning under given settings of the wave plates may enable switching from stationary soliton to breathing soliton states [13-24, 37-39] or to soliton molecules (SMs) with various dynamics [28, 40-42]. Contrary to our previous works [15, 21] and to reduce as much as possible the intracavity loss, we have not implemented any genetic algorithm in the present laser. Indeed, the frequency locking dynamics of breather molecular complexes have been found to be highly affected by cavity damping (see Fig. S1 in the Supplementary Information).

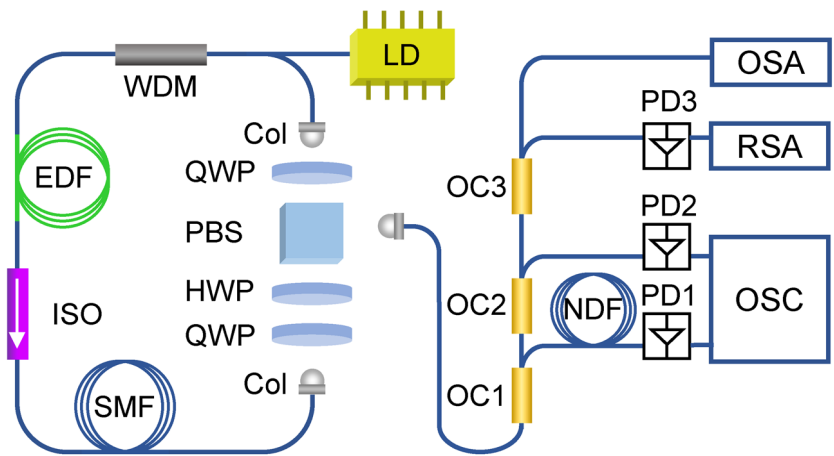

Fig. 1 The laser setup. LD laser diode, WDM wavelength-division multiplexer, EDF erbium-doped fibre, ISO isolator, SMF single-mode fibre, Col collimator, QWP quarter-wave plate, HWP half-wave plate, PBS polarisation beam splitter, OC optical coupler, NDF normally dispersive fibre, PD photodetector, OSC oscilloscope, RSA radio spectrum analyser, OSA optical spectrum analyser.

Figure 2 illustrates how the radiofrequency (RF) spectrum of the laser emission (measured by a radio spectrum analyser connected to photodiode 3 in Fig. 1) is influenced by the pump current (see also the measurements across a frequency range up to $f_r$ provided in Fig. S2 of the Supplementary Information). At low currents (up to 102 mA), our laser emits a single soliton pulse per cavity roundtrip (RT) as evidenced by the single frequency component of the RF signal at $f_r$ (Supplementary Information Fig. S2). Increase of the pump current leads to the generation of a breathing soliton with a short pulsating period of 4 RTs, as revealed by the appearance the subharmonic narrow peaks located at multiples of $f_b = f_r/4$: the breathing frequency $f_b$ is locked to the cavity repetition frequency [19, 21]. The transition from stationary to breathing soliton correlates to the ubiquitous dynamics known as "Hopf bifurcation". Further pump current increase causes new equally spaced spectral lines to appear symmetrically on both sides the subharmonic peaks. The separation $f'_b$ between these new lines, forming a "modulated subharmonic" structure, is associated with a long pulsating period in the time domain. Pump currents above 106 mA break the frequency locking (the modulated sidebands also vanish), and the breathing frequency continuously drifts as the pump strength varies so that $f_b$ is no longer commensurate with $f_r$. We term this laser regime "non-subharmonic breathing". For currents above 111 mA, the pulsating regime disappears, being replaced by a short stage of chaotic-like behavior and then the emission of stationary diatomic SMs up to 117 mA, when pulsating pulses are generated again in the form of breathing SMs. Between 120 and 123 mA, subharmonic and modulated subharmonic behaviors are retrieved whereas non-harmonic features are apparent between 117 and 120 mA and then between 123 and 133 mA. Above 133 mA, the diatomic breathing SM switches to a triatomic molecule, and a similar evolution pattern of the RF spectrum from subharmonic to modulated subharmonic and to non-subharmonic breathing is again recorded.

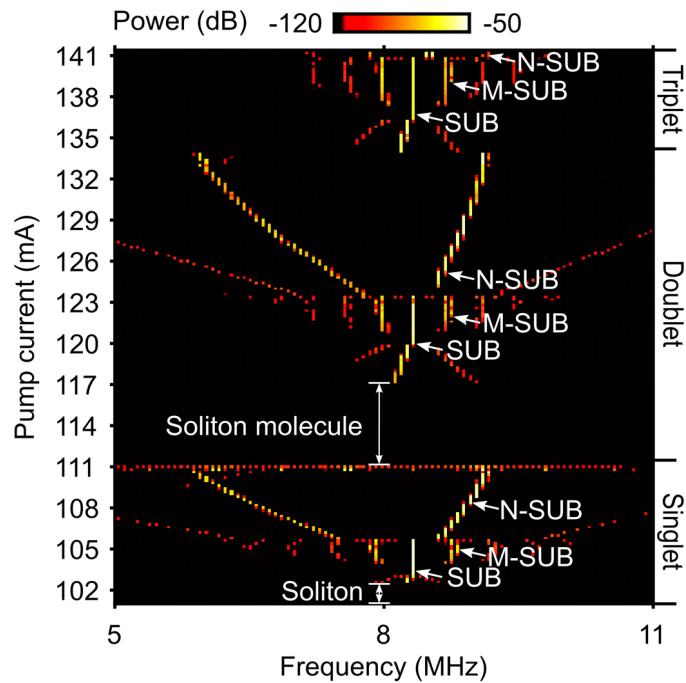

Fig. 2. Map of the laser spectral intensity in the space of RF and pump current, showing phase transitions for single breathers, and diatomic and triatomic breathing SMs. SUB subharmonic, N-SUB non-subharmonic, M-SUB modulated subharmonic.

Our experimental observations emphasise the robustness of the phase transition phenomenon, which is independent of the number of elementary constituents of the breathing soliton structure. To enlighten the laser dynamics in the different phases, we have focused on the case of the diatomic breathing SM that is observed for pump currents in the range 117 to 134 mA. The case of the triatomic breather molecule is illustrated in Fig. S3 of the Supplementary Information, displaying qualitatively similar features. We have complemented the RF spectrum measurements by additional spatio-spectral measurements, which are summarised in Fig. 3. Panels (a-c) therein show the roundtrip-resolved optical spectra measured by the time-stretch technique [43-46] (photodiode 1 in Fig. 1) for the three breathing regimes. Periodic variations of the spectral intensity across a well-defined period of 4 cavity RTs can be observed for the subharmonic regime (Fig. 3a), accompanied by corresponding synchronous periodic changes of the pulse energy (white curve). By contrast, the non-subharmonic regime (Fig. 3b) shows degraded periodicities in both the optical spectrum and energy. The spectra recorded at the RT numbers of maximal and minimal energy within a period (Figs. 3e and 3f) indicate a larger modulation depth for the non-subharmonic state. In both regimes, the period of spectral fringes remains almost unchanged over cavity RTs: the two breathers within the molecule have nearly equal separation, estimated to be 5.5 ps, which does not change over cavity RTs. The modulated subharmonic regime shown in Fig. 3d (close-up of Fig. 3c) features two sets of periodicities with a long period of approximately 88 cavity RTs and a short period of 4 RTs. The details of the corresponding RF spectra in the vicinity of $f_r/4$ shown in Fig. 3g further highlight the differences among the three regimes: the subharmonic state features a single very narrow frequency component located exactly at $f_b=f_r/4$, while a set of equally spaced narrow sidebands on both sides of $f_b=f_r/4$ is observed for the modulated subharmonic case. In sharp contrast to this, the non-subharmonic regime presents much broader spectral lines, confirming the frequency unlocked operation of the laser (see also Fig. S4 in the Supplementary Information for a magnification of Fig. 3g).

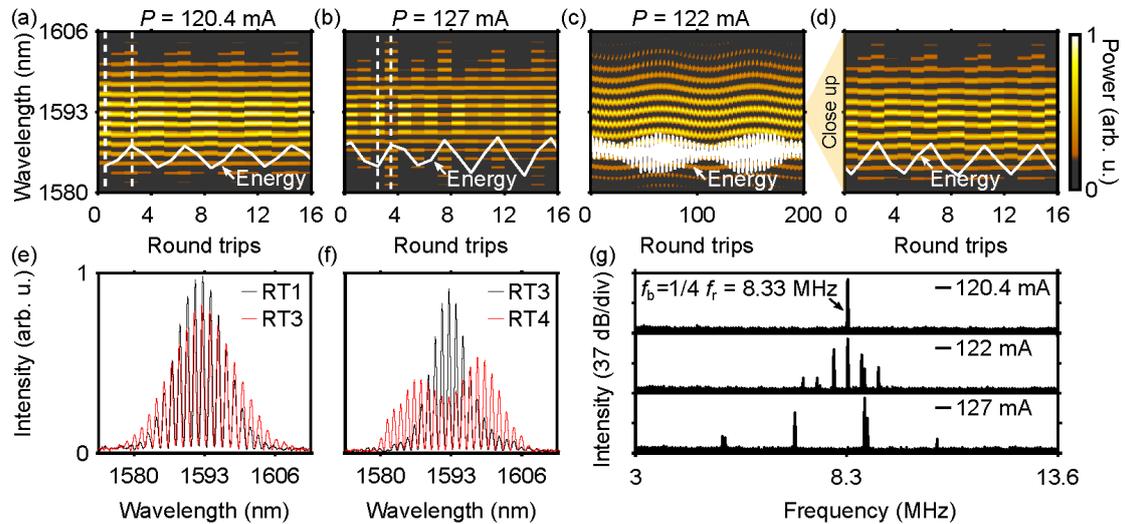

Fig. 3. Experimental observation of subharmonic, modulated subharmonic, and non-subharmonic diatomic breathing SMs. (a-c) Time stretch recording of single-shot optical spectra over consecutive cavity RTs. The white curves denote the energy evolutions. (d) A magnified version of (c) showing the short period breathing. (e,f) Single-shot spectra at the RT numbers of maximal and minimal energies within a period (indicated by dashed lines in (a) and (b), respectively). (g) RF spectrum measurements for the three breathing regimes shown on a reduced frequency span centered on $f_r/4$.

To validate our experimental findings and gain a better insight into the complex temporal dynamics of breathing SMs, we have performed numerical simulations of the laser based on a lumped model which models each part of the laser cavity separately. Pulse propagation in the optical fibres is modelled by a generalised nonlinear Schrödinger equation (NLSE), which includes the effects of GVD and self-phase modulation for all the fibres, and gain saturation and bandwidth-limited gain for the active fibre. In the scalar-field approach, this equation takes the form [21, 28, 30, 47-50]

$$\psi_z = -\frac{i\beta_2}{2}\psi_{tt} + i\gamma|\psi|^2\psi + \frac{g}{2}\left(\psi + \frac{1}{\Omega^2}\psi_{tt}\right), \quad (1)$$

where $\psi = \psi(z,t)$ is the slowly varying electric field moving at the group velocity along the propagation coordinate $z$, and $\beta_2$ and $\gamma$ are the GVD and Kerr nonlinearity coefficients, respectively. The dissipative terms in Eq. (1) represent linear gain as well as a parabolic approximation to the gain profile with the bandwidth $\Omega$. The gain is saturated according to $g(z) = g_0/[1 + E(z)/E_{\text{sat}}]$, where $g_0$ is the small-signal gain, $E(z) = \int dt |\psi|^2$ is the pulse energy, and $E_{\text{sat}}$ is the gain saturation energy. We note that $E_{\text{sat}}$ can be used to change the pulse energy in the numerical simulations, thus playing a similar role to the pump current in the experiment. To match the numerical and experimental pulse energies to the same scale however, rate equations governing the laser gain dynamics should be included into the numerical model. The discrete effects of the effective nonlinear saturation involved in the NPE mode-locking technique are modelled by an instantaneous and monotonous nonlinear transfer function for the field amplitude: $T = \sqrt{1 - q_0 - q_m/[1 + P(t)/P_s]}$, where $q_0$ is the unsaturated loss due to the absorber, $q_m$ is the saturable loss (modulation depth), $P(z,t) = |\psi(z,t)|^2$ is the instantaneous pulse power, and $P_s$ is the saturation power. The numerical model is solved with a standard split-step propagation algorithm and uses similar parameters to the experimental values (Supplementary Table 1).

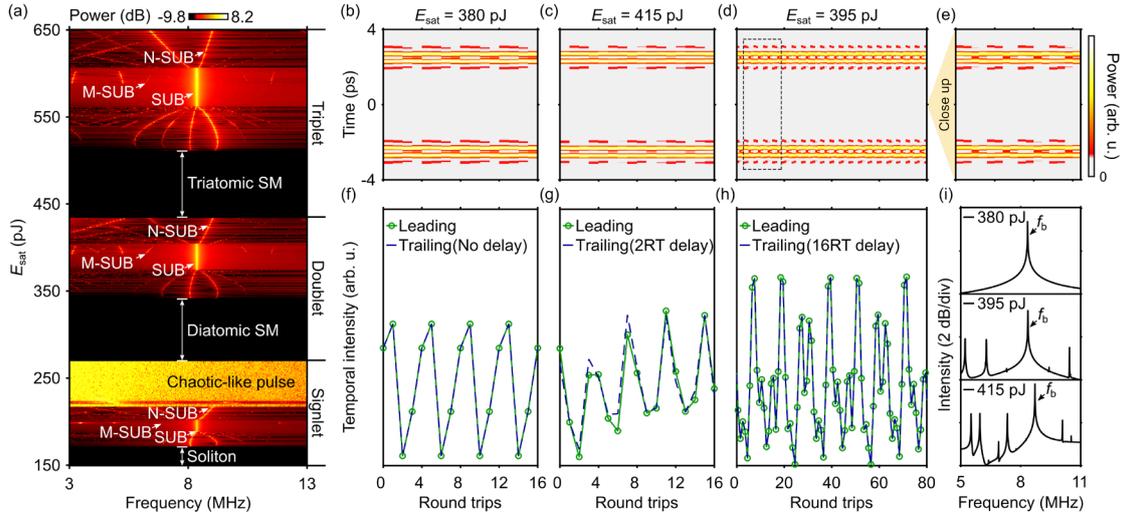

Fig. 4. Numerical modelling results. (a) Transitions among different laser states in the space of RF and gain saturation energy. (b-d) Temporal evolutions of the intensity over consecutive cavity RTs for subharmonic, non-subharmonic and modulated subharmonic diatomic breathing SMs, respectively. Close-up views of the evolutions for the modulated subharmonic breathing SM are shown in panel (e). (f-h) Evolutions of the peak intensities of the leading and trailing breathers for the three breather molecule states. The trailing pulse is delayed by a certain number of RTs in the non-subharmonic and modulated subharmonic states to evidence the synchronisation of the two breathers. (i) RF spectra of the three breather molecule states. Please, change 'chaos-like' to 'chaotic-like'.

The RF spectrum can be obtained by Fourier transformation of the pulse energy in the simulations. The RF spectral intensity as a function of the gain saturation energy plotted in Fig. 4a shows good agreement with the experimental results (Fig. 2) and confirms the existence of an intermediate modulated subharmonic breathing state between the subharmonic and non-subharmonic phases. One may see that compared to the experiments, the simulations tend to overestimate the range of existence of the chaotic-like regime occurring between single and multi-pulse laser emissions. Such a discrepancy can be ascribed to the extremely large parameter space used in the simulations. By increasing the linear intracavity loss in the model or, equivalently, by changing the polarisation state of the laser in the experiments, direct synchronisation-desynchronisation transitions (without the intermediate state) can also be observed in our breather laser (Supplementary Fig. S5). Such direct transitions are saddle-node bifurcations [2]. The experiments (Fig. 2) and simulations (Fig. 4a) also reveal another interesting phenomenon: the higher the number of elementary constituents, the more robust the subharmonic breather structures against pump power (gain saturation energy) variations. Details are given in Fig. S6 of the Supplementary Information.

The roundtrip evolutions of the temporal intensity profiles of subharmonic, non-subharmonic and modulated subharmonic diatomic breathing SMs are shown in panels (b-d) of Fig. 4, respectively. The corresponding spectral evolutions are given in the Supplementary Information (Fig. S7). The evolutions of the peak intensities of the leading and trailing pulses in the subharmonic breathing SM (Fig. 4f) reveal that the two breathers are synchronised to each other. Whilst non-subharmonic breathing SMs are quasi-periodic, there is a constant delay (of 2 cavity RTs) between the two breathers (Fig. 4c), and by adding this delay to the trailing pulse, the evolutions of the two pulses become synchronous (Fig. 4g). This suggests that the dynamics of the two pulses pertains to lag synchronisation, which has been extensively studied in chaotic systems [2, 51-54]. Modulated subharmonic breathing SMs also show lag synchronisation (Fig. 4h). Furthermore, we have also observed lag synchronisation in the dynamics of triatomic breathing SMs. The evolutions of the temporal intensity profiles (Fig. 4b-e) highlight a general feature of the three breathing SM states: each breather within the molecule is split into multiple sub-pulses, which arises from higher-order soliton-like evolution inside the anomalous dispersion segment of the laser cavity (see Fig. S8 and S9 in the Supplementary Information).

In conclusion, we have observed subharmonic and non-subharmonic breathing soliton structures in an ultrafast fibre laser in both experiments and numerical simulations. We have also unveiled the existence of an intermediate state – modulated subharmonic breathing – between the two phases. These findings could stimulate parallel research on the synchronisation of other coupled nonlinear systems where such a state has not been found yet. Furthermore, our results show that transitions among the three states are not restricted to a single-breather laser emission regime, but alike transitions also occur when the laser operates in a diatomic or triatomic breather SM generation regime, which further substantiates the universal nature of synchronisation/desynchronisation phenomena and opens the possibility to study the dynamics of nonlinear systems with three or more interacting frequencies. Mode-locked fibre lasers are very general physical platforms wherein the fibre propagation dynamics are governed by a generalised NLSE. This equation arises in many physical settings including, but not limited to, Bose-Einstein condensates, surface gravity waves, and superconductivity. Therefore, it is reasonable to assume that similar transitions may also arise in these physical systems when a feedback mechanism is included. Following the long history of analogies established between nonlinear fibre lasers and the various states of matter with the establishment of the concepts of molecules, crystals, rains, and gases of optical solitons amongst the others [55], one can also imagine to qualitatively relate the synchronisation/desynchronisation of breathing solitons and breather molecular complexes reported in this paper with commensurate-incommensurate phase transitions [56], which are ubiquitous phenomena in many areas of condensed-matter physics and beyond.

**Acknowledgement**

We acknowledge support from the National Natural Science Fund of China (11621404, 11561121003, 11727812, 61775059, 12074122, 62022033 and 11704123), Shanghai Municipal Science and Technology Major Project (2019SHZDZX01-ZX05), Shanghai Rising-Star Program and Science and Technology Innovation Program of Basic Science Foundation of Shanghai (18JC1412000), the Agence Nationale de la Recherche (ANR-20-CE30-004), the UK Engineering and Physical Sciences Research Council (CREATE – EP/X019241/1), Sustainedly Supported Foundation by National Key Laboratory of Science and Technology on Space Microwave under Grant HTKT2022KL504008, and Shanghai Natural Science Foundation (23ZR1419000).

**References**

1. C. Huygens, *Oeuvres Complètes de Christiaan Huygens*, vol. 5 (The Hague, The Netherlands: Martinus Nijhoff, 1893). Includes correspondence from 1665.
2. A. Pikovsky, M. Rosenblum, and J. Kurths, *Synchronization: A Universal Concept in Nonlinear Science* (Cambridge Universit y Press, Cambridge, UK, 2001).
3. J. K. Jang, A. Klenner, X. Ji, Y. Okawachi, M. Lipson, and A. L. Gaeta, "Synchronization of coupled optical microresonators," Nature Photonics **12**, 688-693 (2018).
4. B. Y. Kim, J. K. Jang, Y. Okawachi, X. Ji, M. Lipson, and A. L. Gaeta, "Synchronization of nonsolitonic Kerr combs," Science Advances **7**, eabi4362 (2021).
5. B. Y. Kim, Y. Okawachi, J. K. Jang, X. Ji, M. Lipson, and A. L. Gaeta, "Coherent combining for high-power Kerr combs," Laser & Photonics Reviews **17**, 2200607 (2023).
6. J. M. Dudley, F. Dias, M. Erkintalo, and G. Genty, "Instabilities, breathers and rogue waves in optics," Nature Photonics **8**, 755-764 (2014).
7. G. Xu, A. Gelash, A. Chabchoub, V. Zakharov, and B. Kibler, "Breather wave molecules," Physical Review Letters **122**, 084101 (2019).
8. F. Leo, L. Gelens, P. Emplit, M. Haelterman, and S. Coen, "Dynamics of one-dimensional Kerr cavity solitons," Optics Express **21**, 9180-9191 (2013).
9. D. C. Cole, and S. B. Papp, "Subharmonic entrainment of Kerr breather solitons," Physical Review Letters **123**, 173904 (2019).
10. E. Lucas, M. Karpov, H. Guo, M. L. Gorodetsky, and T. J. Kippenberg, "Breathing dissipative solitons in optical microresonators," Nature Communications **8**, 736 (2017).
11. M. Yu, J. K. Jang, Y. Okawachi, A. G. Griffith, K. Luke, S. A. Miller, X. Ji, M. Lipson, and A. L. Gaeta, "Breather soliton dynamics in microresonators," Nature Communications **8**, 14569 (2017).
12. J. M. Soto-Crespo, M. Grapinet, P. Grelu, and N. Akhmediev, "Bifurcations and multiple-period soliton pulsations in a passively mode-locked fiber laser," Physical Review E **70**, 066612 (2004).
13. J. Peng, S. Boscolo, Z. Zhao, and H. Zeng, "Breathing dissipative solitons in mode-locked fiber lasers," Science Advances **5**, eaax1110 (2019).
14. T. Xian, L. Zhan, W. Wang, and W. Zhang, "Subharmonic entrainment breather solitons in ultrafast lasers," Physical Review Letters **125**, 163901 (2020).
15. X. Wu, J. Peng, S. Boscolo, Y. Zhang, C. Finot, and H. Zeng, "Intelligent breathing soliton generation in ultrafast fiber lasers," Laser & Photonics Reviews **16**, 2100191 (2022).
16. Y. Du, Z. Xu, and X. Shu, "Spatio-spectral dynamics of the pulsating dissipative solitons in a normal-dispersion fiber laser," Optics Letters **43**, 3602-3605 (2018).
17. M. Liu, Z. W. Wei, H. Li, T. J. Li, A. P. Luo, W. C. Xu, and Z. C. Luo, "Visualizing the "invisible"

soliton pulsation in an ultrafast laser," Laser & Photonics Reviews **14**, 1900317 (2020).
18. K. Krupa, T. M. Kardaś, and Y. Stepanenko, "Real-time observation of double-Hopf bifurcation in an ultrafast all-PM fiber laser," Laser & Photonics Reviews **16**, 2100646 (2022).
19. Z. Wang, A. Coillet, S. Hamdi, Z. Zhang, and P. Grelu, "Spectral pulsations of dissipative solitons in ultrafast fiber lasers: Period doubling and beyond," Laser & Photonics Reviews **17**, 2200298 (2023).
20. Y. Du, Z. He, Q. Gao, H. Zhang, C. Zeng, D. Mao, and J. Zhao, "Emergent phenomena of vector solitons induced by the linear coupling," Laser & Photonics Reviews **17**, 2300076 (2023).
21. X. Wu, Y. Zhang, J. Peng, S. Boscolo, C. Finot, and H. Zeng, "Farey tree and devil's staircase of frequency-locked breathers in ultrafast lasers," Nature Communications **13**, 5784 (2022).
22. J. Peng, Z. Zhao, S. Boscolo, C. Finot, S. Sugavanam, D. V. Churkin, and H. Zeng, "Breather molecular complexes in a passively mode-locked fiber laser," Laser & Photonics Reviews **15**, 2000132 (2021).
23. J. Guo, S. Cundiff, J. Soto-Crespo, and N. Akhmediev, "Concurrent passive mode-locked and self-Q-switched operation in laser systems," Physical Review Letters **126**, 224101 (2021).
24. Y. Cui, Y. Zhang, L. Huang, A. Zhang, Z. Liu, C. Kuang, C. Tao, D. Chen, X. Liu, and B. A. Malomed, "Dichromatic "breather molecules" in a mode-locked fiber laser," Physical Review Letters **130**, 153801 (2023).
25. M. H. Jensen, P. Bak, and T. Bohr, "Complete devil's staircase, fractal dimension, and universality of mode-locking structure in the circle map," Physical Review Letters **50**, 1637 (1983).
26. B. A. Malomed, "Bound solitons in the nonlinear Schrödinger-Ginzburg-Landau equation," Physical Review A **44**, 6954-6957 (1991).
27. M. Stratmann, T. Pagel, and F. Mitschke, "Experimental observation of temporal soliton molecules," Physical Review Letters **95**, 143902 (2005).
28. K. Krupa, K. Nithyanandan, U. Andral, P. Tchofo-Dinda, and P. Grelu, "Real-time observation of internal motion within ultrafast dissipative optical soliton molecules," Physical Review Letters **118**, 243901 (2017).
29. G. Herink, F. Kurtz, B. Jalali, D. R. Solli, and C. Ropers, "Real-time spectral interferometry probes the internal dynamics of femtosecond soliton molecules," Science **356**, 50-54 (2017).
30. J. Peng, and H. Zeng, "Build-up of dissipative optical soliton molecules via diverse soliton interactions," Laser & Photonics Reviews **12**, 1800009 (2018).
31. X. Liu, X. Yao, and Y. Cui, "Real-time observation of the buildup of soliton molecules," Physical Review Letters **121**, 023905 (2018).
32. C. Grebogi, E. Ott, and J. A. Yorke, "Are three-frequency quasiperiodic orbits to be expected in typical nonlinear dynamical systems?," Physical Review Letters **51**, 339 (1983).
33. G. Held, and C. Jeffries, "Quasiperiodic transitions to chaos of instabilities in an electron-hole plasma excited by ac perturbations at one and at two frequencies," Physical Review Letters **56**, 1183 (1986).
34. A. Cumming, and P. S. Linsay, "Quasiperiodicity and chaos in a system with three competing frequencies," Physical Review Letters **60**, 2719 (1988).
35. J. H. Cartwright, D. L. González, and O. Piro, "Universality in three-frequency resonances," Physical Review E **59**, 2902 (1999).
36. D. Noske, N. Pandit, and J. Taylor, "Subpicosecond soliton pulse formation from self-mode-locked erbium fibre laser using intensity dependent polarisation rotation," Electronics Letters **28**,

2185-2186 (1992).
37. J. Peng, and H. Zeng, "Experimental observations of breathing dissipative soliton explosions," Physical Review Applied **12**, 034052 (2019).
38. J. Peng, and H. Zeng, "Triple-state dissipative soliton laser via ultrafast self-parametric amplification," Physical Review Applied **11**, 044068 (2019).
39. Y. Zhou, Y.-X. Ren, J. Shi, and K. K. Wong, "Breathing dissipative soliton molecule switching in a bidirectional mode-locked fiber laser," Advanced Photonics Research **3**, 2100318 (2022).
40. S. Hamdi, A. Coillet, B. Cluzel, P. Grelu, and P. Colman, "Superlocalization reveals long-range synchronization of vibrating soliton molecules," Physical Review Letters **128**, 213902 (2022).
41. D. Zou, Y. Song, O. Gat, M. Hu, and P. Grelu, "Synchronization of the internal dynamics of optical soliton molecules," Optica **9**, 1307-1313 (2022).
42. S. Liu, Y. Cui, E. Karimi, and B. A. Malomed, "On-demand harnessing of photonic soliton molecules," Optica **9**, 240-250 (2022).
43. K. Goda, and B. Jalali, "Dispersive Fourier transformation for fast continuous single-shot measurements," Nature Photonics **7**, 102-112 (2013).
44. A. F. J. Runge, C. Aguergaray, N. G. R. Broderick, and M. Erkintalo, "Coherence and shot-to-shot spectral fluctuations in noise-like ultrafast fiber lasers," Optics Letters **38**, 4327-4330 (2013).
45. A. Mahjoubfar, D. V. Churkin, S. Barland, N. Broderick, S. K. Turitsyn, and B. Jalali, "Time stretch and its applications," Nature Photonics **11**, 341-351 (2017).
46. T. Godin, L. Sader, A. Khodadad Kashi, P.-H. Hanzard, A. Hideur, D. J. Moss, R. Morandotti, G. Genty, J. M. Dudley, and A. Pasquazi, "Recent advances on time-stretch dispersive Fourier transform and its applications," Advances in Physics: X **7**, 2067487 (2022).
47. H. A. Haus, "Mode-locking of lasers," IEEE Journal of Selected Topics in Quantum Electronics **6**, 1173-1185 (2000).
48. D. Mao, H. Wang, H. Zhang, C. Zeng, Y. Du, Z. He, Z. Sun, and J. Zhao, "Synchronized multi-wavelength soliton fiber laser via intracavity group delay modulation," Nature Communications **12**, 1-8 (2021).
49. X. Liu, D. Popa, and N. Akhmediev, "Revealing the transition dynamics from Q switching to mode locking in a soliton laser," Physical Review Letters **123**, 093901 (2019).
50. J. Peng, M. Sorokina, S. Sugavanam, N. Tarasov, D. V. Churkin, S. K. Turitsyn, and H. Zeng, "Real-time observation of dissipative soliton formation in nonlinear polarization rotation mode-locked fibre lasers," Communications Physics **1**, 20 (2018).
51. M. G. Rosenblum, A. S. Pikovsky, and J. Kurths, "From phase to lag synchronization in coupled chaotic oscillators," Physical Review Letters **78**, 4193 (1997).
52. O. Sosnovtseva, A. Balanov, T. Vadivasova, V. Astakhov, and E. Mosekilde, "Loss of lag synchronization in coupled chaotic systems," Physical Review E **60**, 6560 (1999).
53. S. Taherion, and Y.-C. Lai, "Observability of lag synchronization of coupled chaotic oscillators," Physical Review E **59**, R6247 (1999).
54. T. Heil, I. Fischer, W. Elsässer, J. Mulet, and C. R. Mirasso, "Chaos synchronization and spontaneous symmetry-breaking in symmetrically delay-coupled semiconductor lasers," Physical Review Letters **86**, 795 (2001).
55. F. Amrani, A. Haboucha, M. Salhi, H. Leblond, A. Komarov, and F. Sanchez, "Dissipative solitons compounds in a fiber laser. Analogy with the states of the matter," Applied Physics B **99**,

107 (2010).
56. P. Bak, "Commensurate phases, incommensurate phases and the devil's staircase," Reports on Progress in Physics **45**, 587 (1982).